\documentclass[11pt]{article}

\usepackage{fullpage}
\usepackage{amsmath}
\usepackage{amssymb}
\usepackage{mathrsfs}
\usepackage{setspace}
\usepackage{bbm}
\usepackage{dsfont}

\usepackage{graphics}

\usepackage[latin1]{inputenc}
\usepackage[pdftex]{graphicx}

\usepackage{color}
\usepackage[american]{babel}
\usepackage[font=small,labelfont=bf,]{caption}

 \def\p{\partial}
 \newcommand{\bea}{\begin{eqnarray}}
\newcommand{\eea}{\end{eqnarray}}
\newcommand{\be}{\begin{equation}}
\newcommand{\ee}{\end{equation}}
\newcommand{\ba}{\begin{align}}
\newcommand{\ea}{\end{align}}

\newcommand{\RR}{\mathbb{R}} % Reali
\newcommand{\ZZ}{\mathbb{Z}} % Interi
\newcommand{\TT}{\mathbb{T}}

\newcommand{\CW}{\mathcal{W}}

\newcommand{\bn}{{\bf n}}
\newcommand{\bw}{{\bf w}}
\newcommand{\bpsi}{{\bf \psi}}
\newcommand{\bchi}{{\bf \chi}}

\newcommand{\bX}{ {\bf X}}
\newcommand{\bz}{\bar z}
\newcommand{\btau}{\bar \tau}
\newcommand{\W}{{\cal W}}

\begin{document}
\numberwithin{equation}{section}
{
\begin{titlepage}
\begin{center}

\hfill \\
\hfill \\
\vskip -0.45in

\hskip4in IPMU-12-0175, NSF-KITP-12-164
\vskip 0.45in
{\Large \bf Smoothed Transitions in Higher Spin AdS Gravity}\\

\vskip 0.4in

{Shamik Banerjee,${}^a$ Alejandra Castro,${}^{c}$ Simeon Hellerman,${}^d$ Eliot Hijano,${}^{b,e}$ \vskip .5mm  Arnaud Lepage-Jutier,${}^b$ Alexander Maloney${}^{b}$ and Stephen Shenker${}^a$}\\

\vskip 0.3in

{\it $^a$  Stanford Institute for Theoretical Physics and Department
of Physics,Stanford University, CA 94305-4060, USA}\vskip .5mm
{\it$^b$ Physics Department, McGill University, 3600 rue University,
Montreal, QC H3A 2T8, Canada}  \vskip .5mm {\it $^c$Center for the
Fundamental Laws of Nature, Harvard University, Cambridge 02138, MA
USA}\vskip .5mm {\it $^d$ Kavli IPMU, The University of Tokyo
Kashiwa, Chiba 277-8583, Japan}\vskip .5mm {\it $^e$ Department of Physics and
Astronomy, University of California, Los Angeles, CA 90095, USA}
\vskip 0.3in

\end{center}

\vskip 0.35in

\begin{abstract}
\noindent
We consider CFTs conjectured to be dual to higher spin theories of gravity in AdS${}_3$ and AdS${}_4$.  Two dimensional CFTs with $\CW_N$ symmetry are considered in the $\lambda=0$ ($k \to \infty$) limit where they are conjectured to be described by continuous orbifolds.  The torus partition function is computed, using reasonable assumptions, and  equals that of a free field theory.  We find no phase transition at temperatures of order one; the usual Hawking-Page phase transition is removed by the highly degenerate light states associated with conical defect states in the bulk.  Three dimensional Chern-Simons Matter CFTs with vector-like matter are considered on $T^3$, where the dynamics is described by an effective theory for the eigenvalues of the holonomies.  Likewise, we find no evidence for a Hawking-Page phase transition at large level $k$. \end{abstract}

\vfill

\noindent \today

\end{titlepage}
}

\newpage
%\tableofcontents

\section{Introduction}

%Higher spin holography provides a  simple framework to  quantify with accuracy both sides of the duality

Higher spin holography provides a simple and elegant framework to probe our understanding of quantum gravity.  In four dimensions, the simplest version of this duality relates the singlet sector of the three dimensional $O(N)$ model with Vasiliev's higher spin theory of gravity in AdS${}_4$ \cite{Sezgin:2002rt,Klebanov:2002ja, Giombi:2012he}.  The singlet constraint is implemented by coupling to a Chern-Simons gauge theory with large level \cite{Giombi:2011kc}.  In lower dimensions, minimal conformal field theories with $\W_N$ symmetry are related to similar higher spin theories of gravity in AdS${}_3$ \cite{Gaberdiel:2010pz,Gaberdiel:2012uj}.  In both cases, the relative simplicity of the duality makes possible a variety of precise checks.  The boundary theories are potentially exactly solvable due to the presence of an infinite number of conserved charges.  Likewise, the gravity theories -- while complicated -- appear simpler than full string theory in AdS.

Perhaps the most notable feature of these dualities is the apparent paucity of bulk degrees of freedom.   There are higher spin fields present but no conventional strings.  But these higher spin theories contain more than higher spin fields.  They contain so-called ``light states"  \cite{Gaberdiel:2011zw,Chang:2011mz,   Papadodimas:2011pf, Banerjee:2012gh}. In the AdS$_3$ context these correspond to twisted states in a continuous orbifold description of the boundary theory \cite{Gaberdiel:2011aa} as we will review below. They have a bulk interpretation as ``conical defects"  \cite{Castro:2011iw}.   In the AdS$_4$ context, when the spatial boundary is $T^2$,  these states correspond to Chern-Simons holonomies interacting with the vector-like matter \cite{Banerjee:2012gh}.   In the gravitational dual theory, these
are likely described by a topological closed string sector, while the Vasiliev degrees of freedom form an open string sector  \cite{Aharony:2011jz, Chang:2012kt, AHJVFuture}.

The goal of this paper is to understand the effect of these light states on the thermal behaviour of these theories.   It is dramatic.  We will argue that these  states smooth out the Hawking-Page transition, at least in a limit.

 To study such thermal phenomena in the boundary field theory, we put the theory on Euclidean spaces with a compact time direction.   In particular we will compute the partition function of $\W_N$ minimal models on $S^1 \times S^1 = T^2$ in the limit of small 't Hooft coupling $\lambda$.  We will likewise study the partition function of three dimensional Chern-Simons theories with vector matter on $T^2 \times S^1$ in the limit of large Chern-Simons level.

The explicit computations rely on the conjecture that for $\lambda\to0$ the $\W_N$ minimal model is described by a continuous orbifold  \cite{Gaberdiel:2011aa}.\footnote{Further evidence for this conjecture is given in \cite{Fredenhagen:2010zh, Fredenhagen:2012rb, Fredenhagen:2012bw,  Runkel:2001ng}.} This gives a rather simple interpretation of the partition function as a gaussian path integral which can be computed exactly, albeit with some subtleties in the integration measure.  In section \ref{sec:free2} we will show that the partition function equals that of $N-1$ free bosons. To corroborate the conjecture and  justify our assumptions we show that the low dimension part of the continuous spectrum of this free theory matches with the light states of the $\W_N$ CFT as $\lambda \to 0$.

 Our analysis shows explicitly that for the continuous orbifold theory the partition function (and its derivatives) is a smooth function of temperature.  From the bulk gravity point of view this is a surprise.  The AdS$_3$ theory has black hole solutions \cite{Didenko:2006zd,Gutperle:2011kf,Ammon:2012}.  The entropy of the boundary $\W_N$ theory agrees with the Bekenstein-Hawking entropy of these black holes at sufficiently high temperature \cite{Kraus:2011ds,Gaberdiel:2012yb}.
In AdS gravity, the formation of such black holes is typically associated with a Hawking-Page phase transition \cite{Hawking:1982dh}.  During this phase transition the theory jumps from a thermal gas of perturbative states in AdS (with entropy of order 1) to a black hole phase (with entropy of order $N$).
In the dual gauge theory, this is interpreted as a transition between a confined phase at low temperature and a deconfined phase at high temperature \cite{Witten:1998zw}.
The $\W_N$ theory, on the other hand, appears to be in a deconfined phase, with entropy of order $N$, at all temperatures.

 Unfortunately, it is difficult to extend this to the finite $\lambda$ case.  It is not obvious that the CFT spectrum is analytic with respect to $\lambda$, and therefore we cannot blindly extrapolate our results. Nevertheless, the effect of the light states is universal;  even at finite $\lambda$ the entropy at low temperatures is of order $N$.  Thus there is no conventional Hawking-Page phase transition, even at finite 't Hooft coupling.  This does not rule out the possibility of some other sort of phase transition at finite $\lambda$.

In the $2+1$ dimensional boundary dual to AdS${}_4$  we find a similar story.  On $T^2 \times S^1$ we argue that the absence of a Vandermonde determinant in the measure for holonomies indicates the absence of a finite temperature phase transition, at least for large Chern-Simons level.   Here the holonomies allow non-singlet matter states even at low temperatures, giving an intuition for the smoothing out of the Hawking Page transition.

The bulk interpretation of these results requires a better understanding of both  AdS$_3$ and AdS$_4$ higher spin gravity.  There is no phase transition between an AdS ground state and a black-hole solution.   Instead, there should be a continuous family of solutions that smoothly interpolate between these saddles.\footnote{Indeed, in AdS${}_4$ it is not known if  black holes with appropriate boundary conditions exist, though it may be possible to construct such solutions by giving the Vasiliev connection non-trivial holonomy.  See \cite{Didenko:2009td,Iazeolla:2011cb} for a discussion of AdS$_4$ higher spin black holes.  According to the free energy of the boundary CFT, the Hawking-Page transition on a spatial $S^2$ occurs at a Planckian bulk temperature \cite{Shenker:2011zf}.}  This should be described by an appropriate moduli space of solutions in Vasiliev theory (and in AdS$_4$ its completion).  The light states behave like a quantum mechanical system, rather than a local field theory.  Our understanding of their geometrical and topological nature is incomplete.  In  AdS${}_3$, the light states are classically described by conical defects.  In AdS${}_4$, on the other hand, the light states should be related to a topological closed string sector which couples to the open-string Vasiliev fields \cite{Aharony:2011jz, Chang:2012kt, AHJVFuture}.
%We will return to this in the final section, where we elaborate on the bulk interpretation.

The organization of the paper is as follows. In section \ref{sec:free2}  we  compute the partition function  of the $\CW_N$ minimal models in the $\lambda\to0$ limit. The continuous part of the spectrum, due in part  to light $\CW_N$ primaries, contributes logarithmically to the entropy of the system  but dominates for finite temperature at small enough $\lambda$. In section \ref{sec:grav} we interpret these results in the dual AdS$_3$ higher spin theory.   In section \ref{sec:free3} we discuss the analogous behavior for vector-like CFTs  in three dimensions. We end in section \ref{sec:disc} with a discussion of our results and their implications for the geometrical interpretation of Vasiliev's theory.  In appendix \ref{App:Van} we compute the measure of flat connections used in section \ref{sec:free2}.  In appendix \ref{App:WNVerma} we compute the free energy
 and phase structure of a gas of  $\CW_N$ descendants.

\section{Free theory analysis in (1+1) dimensions}\label{sec:free2}

In this section we will compute the partition function for 2D minimal models with ${\CW}_N$ symmetry and zero 't Hooft coupling. Our goal is to understand the phase structure of the CFT and its implications for the phase space and entropy of the dual gravitational higher spin theory.

${\CW}_N$  minimal models are coset WZW models of the form
\be\label{SU}
{SU(N)_k \times SU(N)_1 \over SU(N)_{k+1}}~.
 \ee
 The central charge is
  \be\label{central}
c = (N-1) \left(1-{N(N+1) \over (k+N)(k+N+1)}\right) ~.%\sim N (1-\lambda^2)
%c= (N-1) \left(1 - \lambda^2 {N+1 \over N +\lambda  }\right)~,
\ee
  The 't Hooft limit is defined by  the large $N$ and $k$ limit, where the 't Hooft coupling
\be
\lambda = {N \over N+k}~,
\ee
is held fixed.  In this limit the central charge becomes large and the theory is expected to be dual to a bulk higher spin theory \cite{Gaberdiel:2010pz}.

When $\lambda\to 0$, i.e. $k\to \infty$ with $N$  fixed so that the central charge approaches  $c=(N-1)$, there is  evidence that the theory is described by a continuous orbifold \cite{Gaberdiel:2011aa} (see also \cite{Fredenhagen:2010zh}).  It is the $SU(N)/\ZZ_N$ orbifold of $N-1$ bosons on the torus $\RR^{N-1}/A_{N-1}$, where $A_{N-1}$ is the $SU(N)$ root lattice. This was shown explicitly for $N=2$ in  \cite{Gaberdiel:2011aa}; here we will proceed under the assumption that the proposal is correct for any value of $N$.\footnote{The analysis in section \ref{sec:spec} provides further evidence that the  coset construction for $\lambda=0$ is indeed a free theory. We will elaborate more on this below.  }

We consider the partition function
\be Z_N(\tau) = {\rm Tr} \, q^{L_0} {\bar q}^{{\bar L}_0},~~~~~q=e^{2 \pi i \tau} ~,\ee
where
\be
\tau = \tau_1+i\tau_2 = {1\over 2\pi} (\theta+i\beta)~,\ee
is a complex linear combination of the angular potential $\theta$ and  inverse temperature $\beta$. Since the CFT is rather simple, it is not difficult to write  the partition function as a path integral.  The partition function is
\be\label{eq:path}
Z_N(\tau) = \int_{UVU^{-1}V^{-1}=1} dU dV Z_{\rm eff} (U,V,\tau)~,
\ee
where $Z_{\rm eff}$ is the partition function of $N-1$ bosons on  $\RR^{N-1}/A_{N-1}$ with boundary conditions twisted by $U$ and $V$ in the space and (Euclidean) time directions, respectively.  The integral is over commuting $SU(N)$ holonomies $U$ and $V$.

As the holonomies commute we can take them to be simultaneously diagonalizable; we denote the eigenvalues $e^{i \psi_j}$ and $e^{i \chi_j}$,  $j=1\dots N-1$, and assemble them into vectors $\bpsi$ and $\bchi$. To compute the measure in  \eqref{eq:path}, we treat the holonomies as $SU(N)$ gauge fields and take the gauge coupling to zero.
In appendix \ref{App:Van} we compute the integration measure in detail using this approach.\footnote{It is not obvious that this prescription for defining the integration measure is unique, and we have no strong evidence of uniqueness. However, our answer is compatible with all the symmetries of the orbifold CFT, and most importantly gives a modular invariant partition function.} The answer does not contain any Vandermonde determinants, so
\be\label{eq:measure}
\int dU dV = \int_{\TT^2} d^{N-1} \bpsi d^{N-1} \bchi ~.
\ee
We have denoted the integration range by $\TT$, the maximal torus  $\TT \subset SU(N)$.
One should in principle be careful about the integration range, as we must take into account both the $\ZZ_N$ factor as well as that of the Weyl group $S_N$ on the eigenvalues.  However, the integrand $Z_{\rm eff}(\tau)$ is invariant under the actions of $\ZZ_N$ as well as the action of the Weyl group.  Thus these  will only contribute an overall $\tau$-independent factor to the integral. We will ignore these and all other $\tau$-independent factors, which can be absorbed into the path integral measure.

We now compute the effective action $Z_{\rm eff}=e^{S}$.  Denote by $\bX\in \RR^{N-1}/A_{N-1}$ the free bosons on the $SU(N)$ torus.  The classical solutions to the equations of motion are labelled by momentum and winding numbers $\bn$ and $\bw$ which take values in the lattice $A_{N-1}$
\bea\label{eq:period}
&&\bX (z+2\pi,\bar z +2\pi) = \bX(z, \bar z) +2\pi \bw ~, \cr
&&\bX (z+2\pi\tau,\bar z +2\pi\bar\tau) = \bX(z, \bar z) +2\pi \bn~,
\eea
and hence
\be
\bX_{\bn, \bw} (z,\bar z) = {1\over 2i \tau_2} \left(\bn (z-\bz) + \bw (\tau \bar z - \btau z)\right)~.
\ee
The classical action of such a solution is
\bea
S_{\bn, \bw} &=& {1\over 2\pi}\int d^2 z~ \p \bX_{\bn,\bw} {\bar \p} \bX_{\bn, \bw} %\cr
={\pi\over \tau_2}\left| \bn-\tau \bw\right|^2~.
\eea
To include the effect of the $SU(N)$ holonomies, recall that the gauge transformations in the maximal torus $\TT \subset SU(N)$ generate translations in $\bX$.  Effectively we are shifting the momentum and winding  in \eqref{eq:period} by
\be
2\pi\bn\to 2\pi \bn+\psi ~,\quad 2\pi\bw \to 2\pi \bw+\chi ~.
\ee
Thus the classical action with twisted boundary conditions is
\be\label{eq:action}
S_{\bn, \bw}(\psi, \chi) = {\pi\over \tau_2}\left| \bn + {\psi \over 2\pi }-\tau \bw -\tau{\chi \over 2\pi} \right|^2~,
\ee
so that
\be\label{eq:Zeff}
Z_{\rm eff} = \sum_{\bn, \bw}  {1\over \sqrt{\det \p {\bar \p}}}
e^{-S_{\bn,\bw}(\psi,\chi)}~.
\ee
Note that the one loop determinant is independent of the choice of classical solution (i.e. $\bn$ and $\bw$) because the path integral is gaussian.  Moreover, up to a $\tau$-independent constant, it is equal to the $N^{\rm th}$ power of the determinant of a  single free boson
\be\label{eq:eta}
{1\over \sqrt{\det\p {\bar \p}}} = {1\over \tau_2^{(N-1)/2} (\eta {\bar \eta})^{N-1}}~,
\ee
and is independent of $\psi$ and $\chi$. Here $\eta(\tau)$ is the Dedekind eta function.

Combining \eqref{eq:measure}, \eqref{eq:Zeff} and \eqref{eq:eta} with \eqref{eq:path} we obtain
\be
Z_N =  {1\over \tau_2^{(N-1)/2} (\eta {\bar \eta})^{N-1}} \int_{\TT^2} d \psi d \chi
\sum_{\bn, \bw}
\exp\left({\pi\over \tau_2}\left| \bn + {\psi \over 2\pi }-\tau \bw -\tau{\chi \over 2\pi} \right|^2\right)~.
\ee
Note that $\chi$ and $ \bw$ only appear in the combination $\chi+2\pi \bw$, and analogously for $\bn$ and $\psi$.  We can combine the integral over $\chi$ with the sum over $\bw$ by shifting the integration variable $\chi\to\chi + 2\pi { \bw}$ so that
\bea
\int_\TT d\chi \sum_{\bw} (\ldots)
&=& \sum_{\bw} \int_{\TT_{\bw}}(\ldots)
\cr
&=& \int_{\RR^{N-1}} d\chi (\ldots)~.
\eea
In the first line $\TT_{\bw}$ denotes the shifted range of integration.  In the second line we have used the sum over $\bw$ to decompactify the range of $\chi$ integration.
We can now perform the same manipulations for the $\bn$ sum and the $\psi$ integral to obtain
\bea
Z_N(\tau) &=&  {1\over \tau_2^{(N-1)/2} (\eta {\bar \eta})^{N-1}} \int_{\RR^{N-1}\times \RR^{N-1}} d \psi d \chi \,  e^{-{1\over 4\pi \tau_2}\left| \psi -\tau \chi  \right|^2}~.
\eea
The integral over $\psi$ and $\chi$ is a finite constant independent of $\tau$.
Hence, the $\tau$ dependent part of the partition function is
\be\label{SPF1}
Z_N(\tau) = {1\over \tau_2^{(N-1)/2} (\eta {\bar \eta})^{N-1}} ~,
\ee
This equals the partition function of $N-1$ decompactified free bosons.

This demonstrates that the spectrum of the  continuous orbifold theory -- and hence the thermodynamics -- equals that of $N-1$ uncompactified bosons.  However, the two theories certainly differ at the level of correlation functions.
Nevertheless, this computation establishes that the theory has no phase transition in the large $N$ limit.

We should mention that the partition function is infinite due to the zero modes of the scalars. However we can systematically divide out this divergence (which is independent of the conformal structure $\tau$), and analyze the finite contribution given by \eqref{SPF1}.  The normalization of equation \eqref{SPF1} can be fixed by comparison with the spectrum of light states of the $\W_N$ minimal model, as we will now see.

\subsection{Spectrum revisited}\label{sec:spec}

We found that the partition function in the  $\lambda\to0$ limit of the $\CW_N$ minimal equals that of $N-1$ uncompactified bosons.  In particular  the power of $\tau_2$ in \eqref{SPF1} reflects the continuum of operators $e^{i {\bf k} \cdot \bX}$ and contributes to the total entropy of the system logarithmically in $T$. More explicitly, in the low temperature and large $N$ regime, we find
\be\label{elog}
\log Z_N \sim  {N\over 24} T^{-1}  + {N\over 2} \log T + \dots~,
\ee
where ``$\dots$'' correspond to subleading contributions in $N$ and $T$; the first term is the contribution from the ground state. The logarithmic term is subdominant at sufficiently low temperatures when $T\ll 1$. If instead we are in the regime $T\lesssim 1$, the continuum part of the spectrum is non-negligible and the contribution to the  entropy scales with $N$.\footnote{{Again, we warn the reader that strictly speaking the entropy of the system is infinite. However, this divergence is independent of temperature. This will become more explicit in the following derivations.}} It is  worth highlighting  that the logarithmic correction to the free energy is the first indication of a departure from the Cardy regime for $T\sim O(1)$; this will be relevant when we discuss the gravitational interpretation of our results.

In our derivation we have not made use of the $\CW_N$ coset construction of the theory.  Thus it is unclear which features will persist at finite $\lambda$.  Moreover, it is not clear that the spectrum is an analytic function of $\lambda$. This leaves room to speculate that the lack of a phase transition in the partition function \eqref{SPF1} is an artifact of the free theory limit rather than a generic feature of the theory.  However, in the following we will relate the continuum part of the spectrum to the corresponding  $\CW_N$ primaries, and provide a simple extension of our results for $\lambda=0$ to the finite case.\footnote{{We do not attempt to give here a rigorous analysis of the spectrum for general $\lambda$. This should be done carefully, and there is indication that it can be done in much more detail than the discussion provided here \cite{GGR}.}} This will also provide an interpretation of the logarithmic growth \eqref{elog} which is in accordance with the results reported in \cite{Banerjee:2012gh}.

In general, states can be labelled (in the Drinfeld-Sokolov representation) by pairs of representations of $SU(N)$, say $(\Lambda_+, \Lambda_-)$ (see e.g. \cite{Gaberdiel:2012uj} for a review). As was discussed  earlier by \cite{Gaberdiel:2011zw,Chang:2011mz,Papadodimas:2011pf}, there is a class of states  with $\Lambda_+=\Lambda_- = \Lambda$ whose weight  is given by
\be\label{light}
h(\Lambda, \Lambda) = {C_2(\Lambda) \over (N+k)(N+k+1)} = {\lambda^2 \over N (N+\lambda)} C_2(\Lambda) ~.
\ee
Here $C_2(\Lambda)$ is the quadratic Casimir of the $SU(N)$ representation $\Lambda$. In terms of Young diagram data, we have
\be
\Lambda_i = r_i - {B\over N}~,\quad\quad B= \sum_i r_i~,
\ee
where $r_i$ is the number of boxes in the $i$-th row of the Young diagram, and the $r_i$ are ordered. The quadratic Casimir is
\be\label{Cas}
C_2(\Lambda)=\sum_i n_i^2 ~,\quad \quad  n_i = r_i +{N+1\over 2}-i - {B\over N}~.
\ee
Note that the set of integers  $(n_i +B/N)$ are distinct and there is no repetition among them. We will therefore take $n_1>n_2>\cdots$. In the following, we will shift $C_2(\Lambda)$ by a constant such that the empty Young tableaux has weight zero.

The distinctive feature of the states $h(\Lambda,\Lambda)$ is that  for finite $B$ (i.e. finite number of boxes) the weight goes to zero as $k$ go to infinity; no other primary states in the spectrum have this feature. The states  $h(\Lambda,\Lambda)$  are the so-called `light states,' which form a continuum near the ground state.\footnote{There are also representations where $\Lambda$ scales with $k$, hence not all states with  $\Lambda_+=\Lambda_- = \Lambda$ are truly `light'.}

The contribution of these light states to the partition function can be approximated as follows. Let $\Delta=h(\Lambda, \Lambda)$ be the dimension of a given state.
We want to consider $\Delta \ll 1$ but fixed as $k \rightarrow \infty$.  Equivalently, we want $\beta$ fixed and large as $k \rightarrow \infty.$  Thus we want to study $C_2 \lesssim k^2$, so that $n_i \lesssim k$.   

Next, we can treat (roughly) the $n_i$ as equal to $r_i$ with some offset as $k \gg N\gg1.$   The offset  $B\over N$ acts like a center of mass coordinate for the $r_i$, and the $n_i$ can be thought as relative coordinates.  The $n_i$ add up to zero, but this is just one constraint on $N$ variables, so we will ignore it.  Integrating over the center of mass gives an extra volume factor which will be of order $k$ which we ignore also.
The range of $n_i$ (which can be positive or negative) is of order $k$ and is large compared to the relevant scale determined by $\Delta$, which is $\sqrt{\Delta} k$.  %Therefore we will treat the limits on $n_i$ to be infinite.
Still there is  an ordering on the $n_i$ inherited from \eqref{Cas}.  The partition function becomes
\be
Z=\sum_{n_i = -k \atop n_i {\rm ~ordered}}^k \exp\left(-{\beta\over k^2}\sum_i n_i^2\right)~.
\ee
Define
\be\label{contvar}
x_i={n_i\over k}~.
\ee
Then we can take a continuum approximation as $k\to \infty$
\bea\label{Sc}
Z&=&k^{N-1} \int_{-1}^{1} dx_1 \int dx_2 \ldots \int_{ x_i \, {\rm ordered}}dx_{N-1} \exp\left(-{\beta} \sum_i x_i^2\right)\cr
&=&{k^{N-1} \over (N-1)!} \int_{-1}^{1} dx_1 \int dx_2 \ldots \int dx_{N-1} \exp\left(-{\beta} \sum_i x_i^2\right)
\eea
In the second line we removed the ordering by taking the unordered integral and dividing by $(N-1)!$.
From here it follows that at low temperatures ($\beta$ large) and large $N$ we obtain
\be\label{estimate}
Z \sim {1 \over N!} (k^2 T)^{N/2}~.
\ee
We emphasize that the power law behavior will breakdown at some $T\lesssim 1$. The integrals in \eqref{Sc} are well approximated by gaussians only for $\beta$ large when compared to the volume spanned by the $x_i$'s .  This will be important when we compare our estimate in \eqref{estimate} to numerical data.

The entropy  attributed to the light states (excluding the ground state contribution) is then 
\be\label{ttt}
S \sim  \log Z _{\rm light}= {N\over 2} \log(k^2 T) - N \log N = {N \over 2}\log(T/\lambda^2)~,
\ee
in the large $N$ limit. Therefore, according to \eqref{Sc} and \eqref{ttt}, the density of light states as a function of energy is well approximated by $\rho(\Delta)\sim \Delta^{(N-3)/2}$ for small values of $\Delta$.  (Here we have given the correct finite $N$ expression.)  This matches precisely the density of states of a $N-1$ uncompactified bosons, so provides a non-trivial check of the continuous orbifold result \eqref{SPF1}.

We can check this analysis by directly computing the spectrum of light states for small  $N$ and large $k$.  For $N=2$ (Virasoro unitary minimal models) this can be done analytically.  The primary states are labelled by pairs of integers $(r,s)$ with (see e.g. \cite{DiFrancesco:1997nk})
\be
h(r,s) = {[(k+3)r - (k+2)s]^2-1 \over 4 (k+2) (k+3)}~~~~~~1 \le s \le r < k+2~.
\ee
The light states  \eqref{light} are those with $r=s$.  It is easy to check that as $k\to \infty$ the density of these states behaves like $\rho(\Delta)\sim \Delta^{-1/2}$ for small $\Delta$.  It is important to note, however, that the light states with $r=s$ correctly reproduce this expected density of states only up to $\Delta=1/4$.  In order to reconstruct the continuous orbifold result $\rho(\Delta) \sim \Delta^{-1/2}$ for $\Delta \ge 1/4$ one has to include ``non-light" states. For example, the states with  $r=s+1$ have dimension $\Delta\geq 1/4$ and also contribute to the continuous spectrum.

\begin{figure}
%\centering
{%\includegraphics[width=0.5\textwidth]{N2k1000000.eps}
  \includegraphics[width=0.33\textwidth]{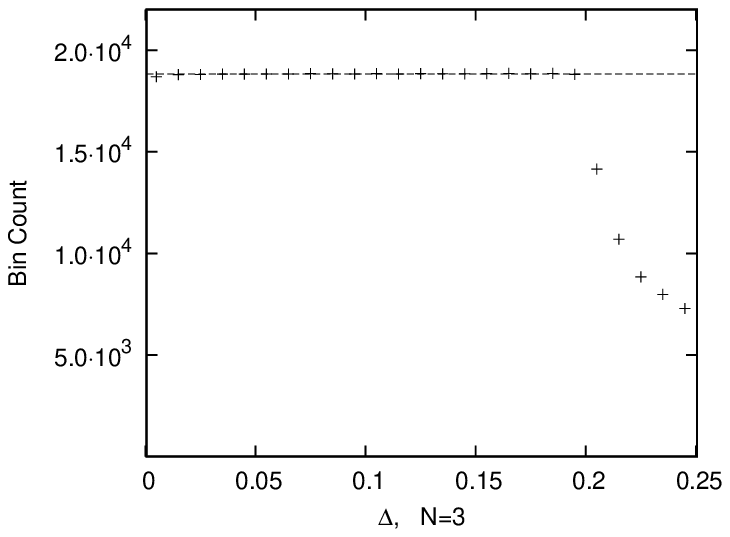}
   \includegraphics[width=0.33\textwidth]{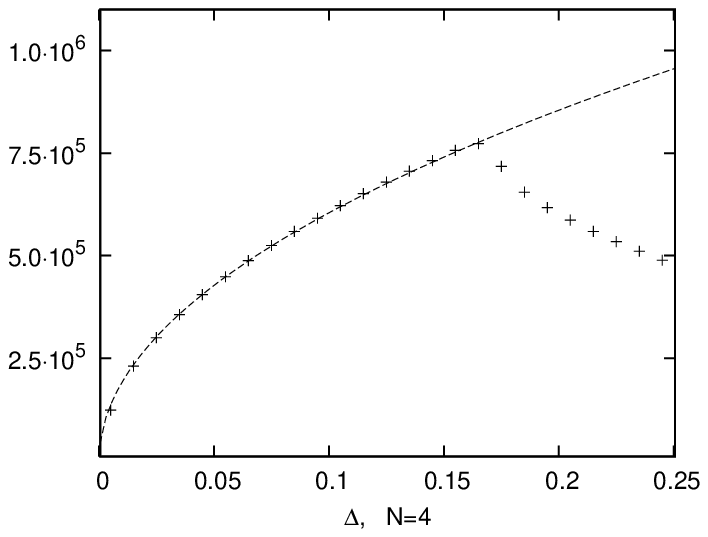}
  \includegraphics[width=0.33\textwidth]{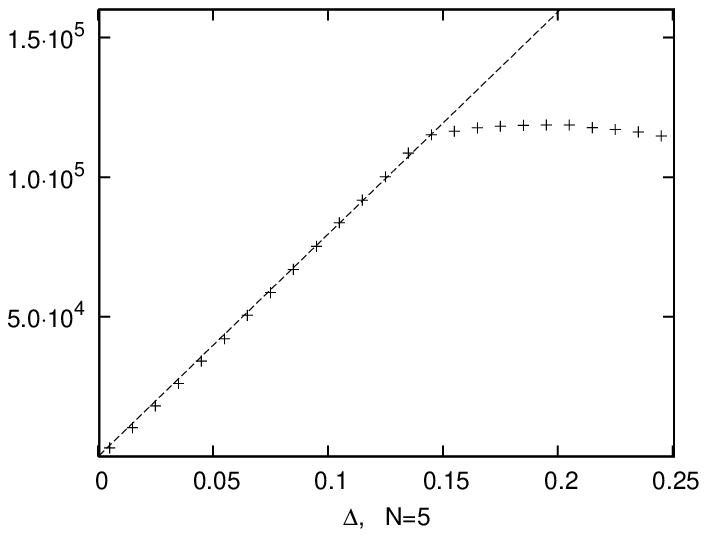}
 }
  \caption{The density of light states $\rho(\Delta)$ for large $k$, denoted by plusses. From left to right we have  $(N=3,k=10^3)$, $(N=4,k=500)$ and $(N=5,k=100)$. The dashed line corresponds to the fit $\rho(\Delta)\sim \Delta^{(N-3)/2}$.  The sharp deviation at large $\Delta$ indicates that, above a critical value of $\Delta$, non-light states give important contributions to the spectrum.}\label{num1}
\end{figure}

\begin{figure}
%\centering
{ \includegraphics[width=0.5\textwidth]{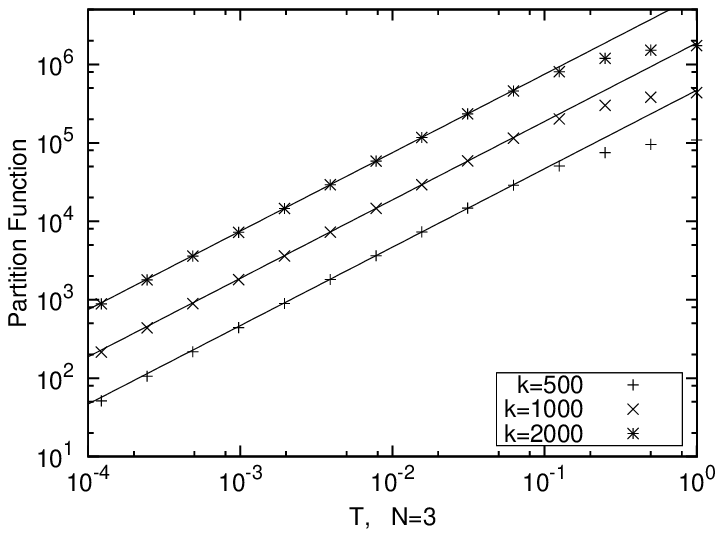}
  \includegraphics[width=0.5\textwidth]{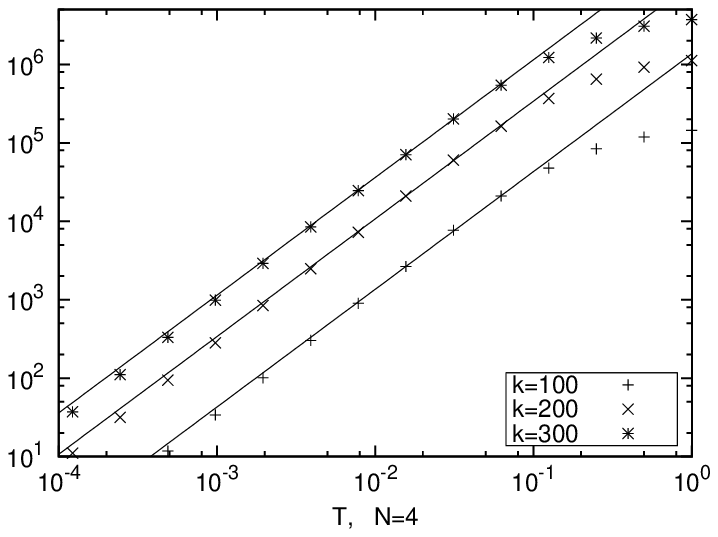}
  } \caption{ $Z$ as a function of $T$ for  $N=3,4$. Scale on both axis are logarithmic. The straight line is the fit $\log Z= (N-1)/2\log T+{\rm constant}$.    }\label{num3}
\end{figure}

In order to extend this discussion to $N>2$ we numerically computed $\rho(\Delta)$ for the light states. We evaluated $C_2(\Lambda)$ for all allowed representations for  $N=3,4,5$ and large values of $k$. In Figs. \ref{num1} and \ref{num3} we display our results.  The density of light states perfectly matches the continuous orbifold result $\rho(\Delta)\sim \Delta^{(N-3)/2}$  for small values of $\Delta$.  Above a finite value of $\Delta$ the light states fail to match the expected density of states. The critical value of the dimension -- the $N>2$ analogue of the $\Delta = 1/4$ point for the Virasoro minimal models -- can be estimated numerically from Fig \ref{num1}.   
%The sharp deviation from the monotonic growth in the numerical computations is due to 
%the breakdown of the saddle point approximation used to obtain \eqref{estimate}.    The distribution of light states is only monotonic for those representations with $\Lambda\lesssim k$, where effectively the counting of states is not sensitive to bound on $r_i$ by $k$. As one approaches $\Lambda \sim k$ the possible representations are sparse and the continuum approximation breaks down.

%\begin{figure}
%\centering
%{
  %} \caption{ Density of light states for  $(N=4,k=500)$ and $(N=5,k=100)$. The dashed line corresponds to the fit $\rho(\Delta)\sim \Delta^{(N-3)/2}$.  }\label{num2}
%\end{figure}

 This is not a surprise.  Just as in the $N=2$ case,  we expect that states with $\Lambda_+ \neq \Lambda_-$ will give important contributions to the spectrum above a certain critical value of the dimension.  
Without attempting a rigorous study of these non-light states, it is easy to see why they should give additional continuous contributions to the spectrum in the $k\to \infty$ limit.  In the language of highest weight representations  the dimension of a state can be written as
\be\label{allstates}
h(\Lambda_+,\Lambda_-)={1\over 2p(p+1)}\left(|(p+1)(\Lambda_++\hat\rho)-p(\Lambda_-+\hat\rho)|^2-\hat\rho^2\right)~,
\ee
where $\hat \rho$ is the Weyl vector of $su(N)$. Here $p\equiv N+k$ and we have added an overall normalization constant so that $h(0,0)=0$.

We want to study the large $p$ limit of this formula.  Let us consider states with $\Lambda_0\equiv \Lambda_+ - \Lambda_-$ fixed and finite as $k\to \infty$. As in \eqref{contvar}, we define
\be
X = {\Lambda_+\over p}~,
\ee
which will be treated as a continuum variable in the large $k$ limit.  As we take $k\to \infty$ the range of $\Lambda_+$ is roughly bounded by $k$, hence we can view $X$ as a vector whose components are continuous degrees of freedom of order one.  The weight \eqref{allstates} becomes
\bea\label{othercont}
h(X,\Lambda_0)= {1\over 2p(p+1)}\left(|pX+p\Lambda_0+\hat\rho|^2-\hat\rho^2\right)
&\underset{p\to \infty\atop \Lambda_0 \, {\rm fixed} }\rightarrow&{1\over 2} |X+\Lambda_0|^2 + O(p^{-1})~.
\eea
Thus a new continuum of states develops, now built on a ground state of positive dimension.  Indeed, this is exactly the behaviour predicted by Fig. \ref{num1}.  It would be interesting to study in more detail the behaviour of the non-light states, and in particular to provide an analytic derivation of the critical dimension  for the non-light states.   

%Basically for fixed $\Lambda_0$ we have a continuum spectrum. And the discussion (including the validity of the approximations) is analogous to  the light states. The important difference for non-zero $\Lambda_0$ is that the starting and end point of the continuum is shifted.

To summarize, as $k\to \infty$, the light states of the $\CW_N$ minimal models are the source of the power law dependence in $\tau_2$ for $Z_N$ at small temperatures.  Their contribution to the entropy grows like ${N \over 2}\log( T/ \lambda^2)$.  The divergent piece in the limit $\lambda=0$ is the characteristic infinite entropy due to the zero modes of the free theory limit.

\section{Implications for AdS$_3$ Higher Spin Gravity }\label{sec:grav}

The behavior of the entropy \eqref{ttt} is not typical of classical gravitational theories. In this section we will discuss the bulk interpretation of the logarithmic corrections to the entropy. This will have implications for the Hawking-Page phase transition for AdS$_3$ higher spin gravity. Most importantly, they indicate that the Bekenstein bound is violated in the bulk gravity theory.

Before discussing the classical spectrum of the higher spin theory, we will review the basic mechanism of the Hawking-Page phase transition in AdS$_3$ \cite{Hawking:1982dh}. We will discuss which conditions lead to the removal of this first order phase transition and discuss which features of the spectrum are  responsible for this effect. This will provide some guidance for the comparison  with the CFT partition function.

For simplicity, we set to zero the angular potential $\theta$ and all other chemical potentials, and consider the free energy as a function of inverse temperature $\beta$. At any fixed temperature there are two relevant classical saddles:  thermal AdS and the BTZ black hole. This statement is independent of the specific matter content of the theory, as the BTZ black hole is a quotient of AdS$_3$ and hence is guaranteed to be a solution of any gravitational theory in  AdS \cite{Banados:1992wn, Banados:1992gq}. Further, the saddles are modular images of each other, with the thermal AdS saddle at inverse temperature $\beta$ the modular transform of the black hole at $4\pi^2/\beta$.  If there is no exotic matter in the gravitational theory --e.g. scalar hair or higher spin fields--   the classical limit of the gravitational path integral is dominated by the free energy of these two solutions
 \be\label{eq:Zgrav}
Z_{\rm grav}(\beta)= e^{\kappa \beta} + e^{4\pi^2 \kappa/\beta} ~, \quad
\ee
where
\be
c=24 \kappa ={3\ell\over 2G} \gg 1 ~,
\ee
with $\ell$ the AdS radius and $G$ Newton's constant.  For the vector-like holography under consideration, we have $c\sim N$ so that the Planck length scales as $G\sim N^{-1}$. Up to numerical factors, in the large $N$ limit we have
\be\label{eq:free}
F(\beta) = -{1\over \beta} \log Z_{\rm grav}(\beta) \underset{N\to \infty}\to
\left\{\begin{array}{cc}
-N & {\rm for} \quad \beta >2\pi \\
-N{4\pi^2\over \beta^2} & {\rm for} \quad \beta < 2\pi
\end{array}\right. ~.
\ee
At $\beta=2\pi$ there is a discontinuity in the first derivative of $F$.  This is the Hawking-Page phase transition.  We emphasize that in order for this phase transition to occur, the free energy must be dominated by the vacuum contribution  {\it for all} $\beta>2\pi$, i.e.
\be\label{eq:limit}
\lim_{N\to \infty}{1\over N} F(\beta>2\pi) = {\rm constant} ~.
\ee
Note that \eqref{eq:limit} always holds at sufficiently low temperatures, e.g. if we scale temperature as $\beta\sim O(N)$. This follows from the existence of a gap in the spectrum above the ground state. However, the requirement that  \eqref{eq:limit} holds up to temperatures of order one is a highly non-trivial constraint.

We now reconsider this derivation for the higher spin duals to the 't Hooft limit of the $\CW_N$ minimal model. In the bulk we are studying a Vasiliev theory with gauge group $hs[\lambda]$. The theory contains an infinite tower of higher spin fields and one complex scalar with mass $M^2=-1+\lambda$ in AdS units. We will not consider the scalar field in this discussion; the higher spin sector is then described by a pair of $hs[\lambda]$ Chern-Simons theories.  The local fields for the graviton (metric) and higher spin analogues can be constructed by the appropriate contractions of the Chern-Simons connections \cite{Henneaux:2010xg,Campoleoni:2010zq, Gaberdiel:2011wb,Campoleoni:2011hg,Henneaux:2012ny}.

The classical phase space of the higher spin theory is defined as follows. It corresponds to the set of Chern-Simons connections that  satisfy AdS fall-off conditions;  this is analogous to the Brown-Henneaux boundary conditions for pure AdS$_3$ gravity \cite{Brown:1986nw}. The solutions also have to be smooth, i.e. the connection has to be globally well defined. In the Chern-Simons language this naturally translates to having trivial holonomy around contractible cycles; see \cite{Gutperle:2011kf, Ammon:2011nk, Castro:2011fm} for a complete discussion of this condition.

The authors in  \cite{Castro:2011iw} found a family of novel solutions to $SL(N)$ higher spin theories which satisfy this condition.\footnote{Ideally one should perform the analysis of smooth classical solutions directly  for $hs[\lambda]$ rather than in $SL(N)$. Unfortunately this  is a difficult task: $hs[\lambda]$ is an infinite dimensional algebra and there is no practical definition of the holonomy of a connection.} These solutions resemble conical defect geometries in a particular gauge.  The description of these solutions as conical singularities is somewhat artificial since the solutions are -- in every gauge-invariant sense -- smooth, but we will continue to refer to these solutions as conical defects. The conical defects have an interesting feature: they are characterized by a set of $N$ fractions $m_i$  such that
\be
m_i = \hat m _i - {{\hat m}\over N} ~,\quad  \hat m=\sum_i \hat m_i ~,
\ee
  $\hat m_i\in\ZZ$ and further with the constraint that $m_i\neq m_j$ for $i\neq j $. Their energies are given by
\be\label{defect}
E=-{c\over N(N^2-1)} C_2(m) ~,\quad \quad C_2(m)=\sum_i m_i^2~.
\ee
The solutions also carry higher spin charges; see  \cite{Castro:2011iw} for further details.

 $SL(N)$ Chern-Simons theory is a semiclassical theory with large central charge $c$ and fixed $N$. This is not in the regime of validity of the $\CW_N$ minimal models since by construction in the CFT we have $c\leq N-1$. Still, we can interpret the conical deficits solutions in the dual CFT as follows. If we compare the Young tableaux construction of the light states \eqref{light} with the spectrum of conical defects, we can identify $m_i=n_i$ in \eqref{Cas}.  The couplings in front of \eqref{light} and \eqref{defect} seem to disagree. This is an artifact which arises because the two expressions are written in different regimes of validity. Using \eqref{central}, we can rewrite \eqref{light} as
 \be\label{light2}
 h(\Lambda, \Lambda) = {C_2(\Lambda) \over (N+k)(N+k+1)}= {N-1-c\over N(N^2-1)}C_2(\Lambda) ~.
 \ee
The authors of   \cite{Gaberdiel:2012ku} showed that the large $c$ and fixed $N$ limit of the minimal models is mathematically well-defined and can be implemented as an analytic continuation of the couplings.\footnote{The analytic continuation makes $c$ independent of $N$, so in this limit $c$ does not necessarily scale with $N$. Further, the CFT is no longer unitary in this limit; this is evident from the negative sign in \eqref{light2} for $c>N$.}    Implementing this here, it is clear that the large $c$ and fixed $N$ limit of \eqref{light2} exactly gives \eqref{defect}. As shown in \cite{Castro:2011iw},  the higher spin charges of the light states and conical defects match in this limit as well.

To summarize, the bulk theory contains a large set of solutions -- conical defects -- which are in one-to-one correspondence with the light states of the boundary CFT. The computation of the path integral for higher spin gravity matches  the computation of the free energy of the CFT, at least at low temperature.  In particular, the conical defects with $C_2(m)$ finite should condense around thermal AdS in the $hs[\lambda]$ theory, forming a continuum in the spectrum. Hence the path integral around thermal AdS will be corrected, just as in section \ref{sec:spec}. The free energy of the higher spin theory for small temperature is then given by \eqref{ttt}
\be\label{HSfree}
F_{\rm HS}(\beta)= F_{N}(\beta)=-{1\over \beta}\log Z_N (\beta) = -{N\over 24} - {N\over 2}T\log(T/\lambda^2)+\dots~.
\ee
Although we can account in the bulk  for the thermal behaviour of the light states, recall that for $\lambda=0$ there are additional non-light states that contribute to the continuum spectrum, such as those of the form \eqref{othercont}.  It would be interesting to understand the gravity interpretation of these states.

The behaviour \eqref{HSfree}  differs significantly from the free energy of a classical gravitational theory \eqref{eq:free}. It is clear that \eqref{HSfree} does not obey the condition \eqref{eq:limit}; there is no Hawking-Page phase transition. A consequence of this observation is that in higher spin gravity a given classical saddle never dominates the free energy at finite temperature. The condensation of light states smears out the phase transition.  This indicates that it is impossible to attribute the thermodynamic behaviour, and hence the entropy, to any individual saddle; one can not isolate the contribution of  classical solution to the free energy. Of course, if we scale $T$ with $\lambda$ and/or $N$ then the thermal AdS or  BTZ black hole will dominate the free energy, but not in the sharp sense defined by the Hawking-Page phase structure.

A curiosity in this analysis is the apparent violation of the Bekenstein entropy bound. The basic intuition that motivates the holographic principle is that entropy in the bulk theory is proportional to area rather than volume. This is equivalent to a linear dependence of black hole entropy on $T$ in three dimensional gravity.  Our findings contradict  this statement. The conglomeration of classical conical defects contributes a large amount of entropy,  in particular a $\log T$ degeneracy per planck length $G\sim N^{-1}$. If the behaviour is geometrical, the effective size of the gas of conical deficits is not linear with length or temperature: each individual conical deficit in the bulk will have some small entropy proportional to its size, however the ensemble of them gives rise to a continuous spectrum with logarithmic growth that dominates the semi-classical entropy.

These conical defect solutions, and in particular their condensation into a continuous spectrum, appear to be special to higher spin theories. We are not  aware of other Einstein-like theories of gravity which display a similar pattern. The thermodynamic properties of higher spin gravity do not resemble the universal features of general relativity and black hole physics.

However, this logarithmic growth is typical of the spectrum of classical string world sheet solutions in AdS$_3$ \cite{Maldacena:2000hw}; long string states also form a continuous spectrum, mimicking some aspects of our discussion.  The novelty of the higher spin theory is that the density of light states scales with $N$, making them abundant in the semiclassical limit. In contrast, the scalar representations of $SL(2,\RR)$ --which represent these long string solutions-- have a density of states that is independent of the coupling.

\section{Chern-Simons Matter  analysis in (2+1)-dimensions}\label{sec:free3}
There are close similarities between the thermal behavior of the $\CW_N$ theory discussed above and that of $SU(N)$ Chern-Simons theory coupled to fundamental scalar matter on a spatial torus $T^2$.   This system has light states \cite{Banerjee:2012gh} described by a global quantum mechanical system with holonomy degrees of freedom that at low energies is just given by $N$ harmonic oscillators with $\hbar = 1/k$.  Here $k$ is the Chern-Simons level.   The semiclassical partition function is the product of two  integrals  of the form (\ref{Sc}), one for the $N$ momenta and one for the $N$ positions.  These give a low temperature entropy
\be
S = N \log(T/\lambda)~,
\ee
where $\lambda = N/k$ is the `t Hooft coupling.\footnote{Here we have given the result for the critical scalar theory.} This is analogous to (\ref{ttt}) after noting that the gaps here are of order $ \lambda$ rather than  $ \lambda^2$.

To study the finite temperature partition function at large $k$ we can first write the pure Chern-Simons partition function in terms of holonomy eigenvalues and then include the effects of the scalar matter.   On $S^2 \times  S^1$ the Vandermonde determinant causes eigenvalue repulsion while the scalar action favors all eigenvalues to be at the origin.   These effects balance at a Gross-Witten-Wadia phase transition at a temperature of order $\sqrt{N}$ \cite{Shenker:2011zf, Giombi:2011kc}.

On $T^2 \times S^1$ the situation is quite different.  There is no Vandermonde determinant for the eigenvalues in the pure Chern-Simons theory \cite{Blau:1993tv, Beasley} and so no balancing forces on the eigenvalues to cause a phase transition.   So we conclude that at least in the limit of large $k$  there is no phase transition in this system on a spatial $T^2$.  The eigenvalues of the ``thermal" circle are concentrated around the origin and so the system behaves much like one without a singlet constraint imposed. This absence of phase transition is analogous to the $\CW_N$ result discussed above. To extend this argument to large but finite $k$ it will be necessary to study perturbative corrections to the measure on holonomies.   We have not done this, but we do not expect a qualitative change.

An intuitive reason for the difference between spatial $S^2$ and $T^2$ is the following. The Gauss Law constraint is, schematically:
\be
k F_{12}^a = j_0^a~,
\ee
where $F_{12}^a$ is the gauge field strength and $j_0^a$ is the scalar charge.
On $S^2$ the only solution at large $k$ is $j_0^a = 0$, i.e. scalar singlet states.  One has to go to temperatures $T \sim \sqrt{N}$  to make this singlet constraint unimportant \cite{Shenker:2011zf, Giombi:2011kc}.   On $T^2$ however the presence of almost flat connections allows states with $F_{12}^a \sim 1/k$.\footnote{Schematically, $kF_{12}^a  = f^{abc}Q^a P^b$ where $Q^a$ and $P^a$ are canonically conjugate holonomies  \cite{Banerjee:2012gh}.}  These satisfy the Gauss Law constraint with nonzero $j_0^a$.  So non-singlet scalar states are allowed.   The presence of these states allows a smooth evolution to the high temperature $NT^2$ unconstrained scalar entropy.

\section{Discussion}\label{sec:disc}

The geometric interpretation of higher spin gravity remains puzzling.  Among other issues, we must understand the black holes of the theory.  In AdS$_3$ higher spin theories we have a sharp definition of the classical solitons, and thus explicit black hole solutions.  In particular, it is also understood how to describe solutions carrying higher spin charges. A  non-trivial test of vector-like dualities in AdS$_3$/CFT$_2$ is the exact agreement between the thermodynamics of the higher spin black hole and the high temperature spectrum of the CFT  \cite{Kraus:2011ds,Gaberdiel:2012yb}.  The black hole carries higher spin charge, thus additional sources are turned on in the CFT partition function. The high $T$ density of states is determined by the modular properties of the partition function with arbitrary number of insertions of the zero mode of the higher spin charge.

However, the theory remains enigmatic. In particular, other (apparently non-black hole) gravitational solutions are important. For instance, in AdS$_3$ higher spin gravity there are more smooth classical solitons -- the conical defects-- than is usually the case in gravitational theories.  These solutions are crucial for the $\CW_N$ minimal model correspondence proposed by  \cite{Gaberdiel:2010pz}, as these conical solutions are in one-to-one correspondence with the light primary states of the dual theory \cite{Castro:2011iw,Gaberdiel:2012ku}.  We have explored the effects of this light sector on the thermodynamics of the system.  Our results reveal a novel feature of the bulk states, that they smooth out the Hawking-Page phase transition. It would be interesting to study the higher genus partition function of $\W_N$ CFT, and investigate whether the higher genus version of the Hawking-Page transition is smoothed out as well.

To understand the disappearance of the Hawking-Page transition in AdS$_3$/CFT$_2$,  we evaluated the partition function  at $\lambda=0$ using the description of the WZW coset \eqref{SU} as a continuous orbifold CFT.  We emphasize that this is a conjecture; there is no proof that the $\CW_N$ theory reduces to this specific free theory at zero coupling. Our analysis in section \ref{sec:spec} gives further evidence for this conjecture. We have matched the spectrum of the light states to the continuous spectrum of the free theory.

The  consequences of this smoothed transition are significant.  The light states have an entropy of order $\log(T)$, rather than $T$.  This indicates that these states are quantum mechanical rather than  field theoretic. They do not correspond to ``local" degrees of freedom.  In the AdS${}_4$ case, these states appear to be topological in nature. %and the same may be true in AdS${}_3$.
 In the continuous orbifold theory, the light states effectively behave as  the zero mode of the $N-1$ uncompactified bosons, which is the quantum mechanics of the continuum of operators $e^{i{\bf k}\cdot \bX}$ in the free field theory language.

The presence of a continuous spectrum is novel but not disturbing. The shocking news is that  there are a lot of them.   In fact the entropy of these states dominates the entropy until a temperature where $N \log(T/\lambda^2) \sim N T$.  This gives a scale
\be\label{scaleT}
T_0 \sim \log(1/\lambda^2) + \ldots ~,
\ee
at small $\lambda$.    It will be important to understand the physical meaning of this scale.  One possible approach would be to examine the bulk conical deficit solutions in the presence of a BTZ black hole.  In a particular there has to be a dynamical mechanism that predicts the crossover in the bulk. Since the scale \eqref{scaleT} is governed by the 't Hooft coupling, which also fixes  the mass of the scalar in Vasiliev's theory, it might be necessary to include backreactions from the scalar field in the analysis. This field also adds local degrees of freedom -- recall that $hs[\lambda]$ Chern-Simons is topological -- and hence might be a natural mechanism for the ``gravitational collapse'' of a gas of conical deficits into a black hole. We emphasize that these last remarks are highly speculative, and to address this properly we must construct a representation of the conical solutions in the $hs[\lambda]$ theory.
% [To make this analysis geometrical do we need to work in a gauge where the conical deficits are wormholes?  Or are the conical deficits inherently nongeometrical and the black holes geometrical?]

The situation in AdS$_4$/CFT$_3$ is somewhat similar.  The partition function on $T^2 \times S^1$ has a similar crossover and hence no phase transition.   The entropy of the lights states is $N \log(T/\lambda)$.  The unconstrained scalar field (or conjectural black hole) entropy is $N T^2$.  So the crossover is at $T_0 \sim \sqrt{\log(1/\lambda)} $.  In the case of $\Sigma_g \times S^1$ the scale is even higher.  Here the entropy of the light states is $N^2 \log(T/\lambda)$ so the crossover temperature is $T_0 \sim \sqrt{N \log(1/\lambda)}$.

The bulk interpretation here is likely related to the observation that the light states found in \cite{Banerjee:2012gh} are described by a topological closed string sector  of  an open-closed string theory where the Vasiliev excitations are open string states \cite{Aharony:2011jz, Chang:2012kt, AHJVFuture}.

\section*{Acknowledgements}

{We are grateful to Chris Beasley, Tom Hartman, Shiraz Minwalla,  Wei Song, Arkady Vainshtein, and Xi Yin for useful discussions and to Matthias Gaberdiel, Rajesh Gopakumar and Mukund Rangamani for valuable comments on an earlier version of this paper. We are especially grateful to Matthias Gaberdiel for discussions related to the continuous orbifold conjecture.  In addition we thank the participants of the KITP program ``Bits, Branes and Black holes'' and the ESI workshop on ``Higher Spin Gravity'' for useful discussion.  The research of SB  and SS is supported by NSF grant 0756174 and the Stanford Institute for Theoretical Physics. The work of AC, ALJ  and AM is supported by the National Science and Engineering Research Council of Canada. AC, AM and SS  are supported in part by NSF under Grant No. PHY11-25915. EH acknowledges support from ''Fundaci\'on Caja Madrid".  The work of S.H. was supported by the World Premier International Research Center  Initiative, MEXT, Japan, and also by a Grant-in-Aid for Scientific Research (22740153) from the Japan Society for Promotion of Science (JSPS). The work of AC is also supported by the Fundamental Laws Initiative of the Center for the Fundamental Laws of Nature, Harvard University.
}

\appendix

\section{Measure factor}\label{App:Van}

Our goal is to compute the integration measure of the path integral \eqref{eq:path}. A convenient way to cast the integral over the $SU(N)$ holonomies $U$ and $V$ is by treating them as components of a $SU(N)$ gauge field $A_\mu$ and taking the gauge coupling to zero.  The action for $SU(N)$ gauge fields is
as
\be S={1\over g^2}\int d^{2}x\, {\rm tr}\left[ F_{\mu\nu}F^{\mu\nu}
\right] ~,\ee
where $g$ is the gauge coupling. The discussion follows \cite{Aharony:2005ew};  the only minor difference is the inclusions of matter fields.

We compactify the two dimensions on a torus of radii $R_{\mu}$. We first characterize the zero modes of our
theory, i.e. the modes whose action vanishes. The zero modes for $A_{\mu }$
consist of diagonal matrices with $N$ eigenvalues $\psi _{\mu }^{I}$, i.e.
\be
\begin{split}
A_{\mu }^{IJ} &\rightarrow \psi _{\mu }^{I}\,\delta ^{IJ}+\hat{A}_{\mu
,\left\{ m_{\nu }\right\} }^{IJ}
\end{split} ~,
\ee
for $\left\{ m_{\nu}\right\}$ labelling the modes of the fields around the two cycles of the spacetime torus. Then the action to quadratic order is

\be
\begin{split}
\label{action}
S=&{1\over g^2}\sum\limits_{\left\{
m_{\mu }\right\} }\sum\limits_{I<J}\sum\limits_{\mu \neq \nu}\Bigg\{ \left( \left( \Delta \psi
^{\mu
}\right) _{IJ}-\frac{2\pi m^{\mu }}{R_{\mu }}\right) ^{2}\left\vert \hat{A}
_{\nu,\left\{ m_{\lambda }\right\} }^{IJ}\right\vert ^{2} \cr &
-\left( \left( \Delta \psi
^{\mu }\right) _{IJ}-\frac{2\pi m^{\mu }}{R_{\mu }}\right) \left(
\left( \Delta \psi ^{\nu}\right) _{IJ}-\frac{2\pi m^{\nu}}{R_{\nu}}\right)
\hat{A}_{\mu ,\left\{ m_{\lambda }\right\} }^{IJ}\hat{A}_{\nu,\left\{
m_{\lambda }\right\} }^{IJ\ast}
\Bigg\} ~,
\end{split}
\ee
where $\Delta \psi^{\mu}_{IJ}=\psi^{\mu}_{I}-\psi^{\mu}_{J}$ is the difference of eigenvalues. We now want to integrate out the gauge field KK modes $\hat{A}_{\mu
,\left\{ m_{\lambda }\right\} }^{IJ}$. The naive result will be
$1/\det M$ where $M$ is a $2$ by $2$ matrix that can be written as

\be
\begin{split}
M &=\left( D^{T}D\right) I-DD^{T} ~, \\
D^{T} &=\left( \left( \Delta \psi ^{1}\right) _{IJ}-\frac{2\pi m^{1}}{R_{1}}%
,\left( \Delta \psi ^{2}\right) _{IJ}-\frac{2\pi m^{2}}{R_{2}}\right) ~.
\end{split}
\ee

The eigenvalues of this matrix are $0$ and $\sum\limits_{\mu }\left(
\left(
\Delta \psi ^{\mu }\right) _{IJ}-\frac{2\pi m^{\mu }}{R_{\mu }}\right) ^{2}$%
. The existence of a vanishing eigenvalue is an indicator that we have
forgotten about gauge symmetry. Convenient gauge fixing constraints are
\be
\begin{split}
\partial _{1}A_{1} &=0 ~,\\
\partial _{2}\int dx_{1}A_{2} &=0 ~.
\end{split}
\ee

Up to a normalization, the Faddeev-Popov measure will come from the following determinant
\be
\det{}^{\prime }\left(\partial _{\mu}-i\left[ A_{\mu},\ast \right] \right) ~,
\ee
which evaluates to
\be\label{FP} \prod\limits_{I<J}\prod\limits_{m_{\lambda }}\left(
\left( \Delta \psi ^{2}\right) _{IJ}-\frac{2\pi m^{2}}{R_{2}}\right)
^{2} ~. \ee

The gauge fixing constraints in terms of KK modes read
\be
\begin{split}
A_{\left\{
r,m_{2}\right\}}^{1} &=0 ~,\\
A_{\left\{
0,r\right\}}^{2} &=0 ~,
\end{split}
\ee
where $r\neq 0$. Exactly one component of the gauge field is
eliminated by these constraints, which means that $M$ is now a $1$ by
$1$ matrix with eigenvalue
$\prod\limits_{I<J}\prod\limits_{m_{\lambda }}\left( \left( \Delta
\psi ^{2}\right) _{IJ}-\frac{2\pi m^{2}}{R_{2}}\right) ^{2}$. From
this it is obvious that the Faddeev-Popov determinant \eqref{FP} and the
factor coming out of the integration of KK modes cancel exactly.
There is no Vandermonde determinant.

\section{Gas of Free Higher Spin Particles}\label{App:WNVerma}
 In this appendix we study the thermodynamics of a gas of higher spin particles in AdS$_3$/CFT$_2$. These theories possess both $\CW_N$ primary and $\CW_N$ descendant states; for simplicity we consider possible phase transitions due to descendant states.   These are easiest to study in the case $\lambda=0$, which corresponds to taking $k\to\infty$ first before taking $N\to \infty$.  In this case no null states are removed from the spectrum and the descendant states live in the $\CW_N$ versions of the Verma module.

  We rewrite the $\lambda=0$ partition function \eqref{SPF1} in terms of the $\CW_N$ characters
\be\label{eqZfree}
Z_N = |q|^{-(N-1)/12} \left( {1\over \tau_2^{(N-1)/2}} \left|\prod_{n=1}^{N-1} (1-q^n)^{n-N} \right|^2\right) \chi_N {\bar \chi}_N ~,
\ee
with
\be
\chi_N = \prod_{s=2}^N \prod_{n=s}^\infty (1-q^n) = \prod_{n=1}^{N-1} (1-q^n)^{N-n} P(q)^{N-1} ~,
\ee
and
\be
P(q) = q^{1/24}\eta(\tau)^{-1} ~.
\ee
The prefactor in \eqref{eqZfree} reflects the standard (cylinder) normalization where the ground state has dimension $-c/24$.  The quantity in parenthesis encodes the density of states of $\CW_N$ primaries, and the free  $\CW_N$  character is $\chi_N$.  The partition function
\be
Z(\tau)= |\chi_N(\tau)|^2 ~,
\ee
describes a gas of free higher spin excitations (the analogues of boundary gravitons).

Let us first verify that this gas of higher spin particles reproduces the expected Cardy behaviour.  We will do this by studying the higher temperature ($\beta \to 0$) behaviour of $\chi_N$. The asymptotics of $P(q)$ are easy to determine.  At high temperature
\be
P\sim \exp\left({\pi^2 \over 6} \beta^{-1} + {1\over 2} \log {\beta } - {1\over 2} \log 2 \pi -{\pi \over 24} \beta + \dots \right)  ~.
\ee
It remains only to understand the asymptotics of the polynomial prefactor.
We will define its logarithm to be
\bea
g &=& \log \prod_{n=1}^{N-1} (1-q)^{N-n}
=\sum_{n=1}^{N-1} (N-n) \log (1-q^n)
\cr
&=&-\sum_{m=1}^\infty \sum_{n=1}^{N-1} {N-n\over m} q^{n m}
\cr &=& - \sum_{m=1}^\infty {q^{N m} - N q^m +N-1  \over m(q^m -2 + q^{-m})} ~.
\eea
This can be approximated as $\beta\to 0$, as
\be
g\sim {N(N-1)\over 2} \log \beta + g_0 - {N (N^2 -1) \over 12} \beta +\dots  ~.
\ee
We see that at fixed $N$ and large temperature $(\beta \to 0)$ the $\CW_N$ character is dominated by $P(N)$
\be
\log \chi_N \sim {\pi^2 (N-1) \over 6} \beta^{-1} +\dots ~.
\ee
This agrees with the free energy \eqref{eq:free} in the large $N$ limit.

We now wish to understand the robustness of this result in the large $N$ limit.  We know that for any fixed $N$ the free energy will scale like $\beta^{-2}$ at sufficiently high temperature, but we do not know how large $\beta$ must be taken in order for this result to apply.  It will turn out that $F \sim \beta^{-2}$ only when $\beta$ vanishes more quickly than $N^{-1}$ in the large $N$ limit.  In other words, Cardy's formula is only valid in the regime where the temperatures are large compared to $N$. Moreover we will not find a phase transition as a function of temperature.

To see this, let us reconsider the formula for $g$ above.  We will scale the temperature with $N$ as $\beta = \gamma N^{-p}$ where $\gamma$ is held fixed in the large $N$ limit.  It is straightforward to expand the polynomial appearing in $g$ at large $N$.  Let us first consider the case  $p<1$, where we obtain at large $N$
\be
g \sim  \sum_{m=1}^\infty\left({1\over m^3 \beta^2} - {N \over m^2 \beta} + \dots  \right) ~,
\ee
where the $\dots$ denotes lower powers of $\beta$, which are subleading as $\beta \to 0$. The sum over $m$ gives
\footnote{The ``$\dots$'' terms in \eqref{eqxx} actually contain terms of the form $\sum m^d = \zeta(-d)$, with $d>-1$, which must be regulated using zeta function regularization.  This is because, although $g$ itself is a manifestly convergent sum, we have reorganized it here into a sum of divergent sums.}
\be\label{eqxx}
g\sim {\zeta(3)}\beta^{-2}- {N \pi^2 \over 6}\beta^{-1} + \dots  ~.
\ee
Combining this with the asymptotic expression for $P(q)$ we find that the $\beta^{-1}$ terms cancel.  We are left with
\be
F \sim - {\zeta(3)\over  \beta^3} ~.
\ee

Let us now consider the case where $p>1$.  One can again expand $g$ at large $N$, and in this case it is easy to see that the leading terms at large $N$ are all positive powers of $\beta$.  The free energy is dominated by $P(q)$, and we obtain the usual Cardy behaviour
\be
F \sim -{N \pi^2 \over 3 \beta^2} ~.
\ee

In order to understand the transition between these two regimes, let us consider the case where $p=1$, so that $\beta N$ is held fixed in the large $N$ limit.  In this case
\bea
g&\sim& \sum_{m=1}^\infty \left({1-e^{-m N \beta} \over m^3 \beta^2} -{N \over m^2 \beta}\right)\cr
&  =&{\zeta(3) - Li_3(e^{-N\beta}) \over \beta^2} -{N \pi^2 \over 6} \beta^{-1}+\dots  ~.
\eea
Now, as $\beta N \to 0$ we have
\be
 Li_3(e^{-N \beta}) \sim \zeta(3) - {\pi^2 \over 6} N \beta + \dots ~,
 \ee
so that $g$ approaches a constant as $\beta N \to 0$.  In this case the free energy has the Cardy behaviour.  On the other hand, $Li_3(e^{-\beta N})$ vanishes exponentially when $\beta N$ is large.  This leads to the non-Cardy, $F \sim 1/ (N \beta^3)$ behaviour.   At intermediate values of $\beta N$ the function $Li_3$ is perfectly smooth.  We conclude that when we take the large $N$ limit with $\beta N$ fixed, the free energy -- regarded as a function of $\beta N$ -- smoothly interpolates between the two behaviours described above.

It is worth noting that the behaviour described above does not constitute a phase transition in the usual sense, in that the free energy is a smooth function of $\beta $ for each of the scaling behaviours described above.  However, the free energy is non-analytic in the following sense.  Instead of regarding the free energy $F(\beta, N)$ as a function of $\beta $ and $N$, we can let $\beta = \gamma N^{-p}$ and regard the free energy as a function of $N, \gamma$ and $p$.  Then in the large $N$ limit at fixed $\gamma$ we have
\be
F(N,\gamma,p)\to - \left\{ {\zeta(3)\gamma^{-3} N^{3p}~~~~~p<1 \atop {\pi^2 \over 3} \gamma^{-2} N^{2p+1} ~~~~~p>1}\right. ~.
\ee
Thus, the free energy is not an analytic function of $p$, however it is important to keep in mind that the free energy at fixed $p$ is an analytic function of $\gamma$, i.e. of the temperature.

\end{document}